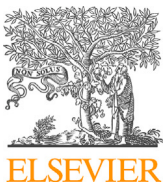
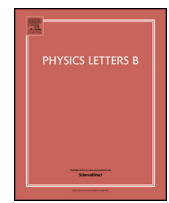

Letter

# Uniform descriptions of pseudospin symmetries in bound and resonant states

Ting-Ting Sun [a,b],[*], Zhi Pan Li [c]

[a] *School of Physics, Zhengzhou University, Zhengzhou 450001, China*
[b] *Guangxi Key Laboratory of Nuclear Physics and Nuclear Technology, Guangxi Normal University, Guilin 541004, China*
[c] *School of Physical Science and Technology, Southwest University, Chongqing 400715, China*

A R T I C L E   I N F O

Editor: A. Schwenk

A B S T R A C T

As a continuation of our previous work on the conservation and breaking of the pseudospin symmetry (PSS) in resonant states [Phys. Lett. B **847**, 138320 (2023)], in this work, the PSS in nuclear single-particle bound and resonant states are investigated uniformly within a relativistic framework by exploring the poles of the Green's function in spherical Woods-Saxon potentials. As the potential depths increase from zero to finite depths, the PS partners evolve from resonant states to bound states. In this progress, the PSS is broken gradually with energy, width, and density splittings. Specially, the energy and width splittings for the resonant and bound states are directly determined by the ratio of the pseudo spin-orbit potentials between the PS partners. Obvious threshold effect is observed for the energy splitting at a critical potential depth, with which one PS partner has become a quasi-bound state inside the centrifugal barrier while the other one is still a high-energy resonant state outside the centrifugal barrier. The differences in the density distributions of the lower component between the PS partners are manifested in the phase shift for the resonant states and amplitudes for bound states.

## 1. Introduction

Symmetries in the single-particle spectrum of atomic nuclei are of great importance on nuclear structures and have been extensively studied in the literature (see Refs. [1,2] and references therein). Pseudospin symmetry (PSS), i.e., the two single-particle states with quantum numbers $(n, l, j = l + 1/2)$ and $(n - 1, l + 2, j = l + 3/2)$ are quasi-degeneracy, was found in atomic nuclei more than 50 years ago, and can be redefined as the pseudospin (PS) doublets $(\tilde{n} = n, \tilde{l} = l + 1, j = \tilde{l} \pm 1/2)$ [3,4]. Afterwards, PSS has been used to explain a number of phenomena in nuclear structures, such as deformation [5], superdeformation [6,7], identical rotational bands [8,9], magnetic moment [10], quantized alignment [11] and so on. Besides, in atomic and molecular physics, PSS has also attracted great concerns and has been discussed in some special atomic and molecular potentials [12–14].

Until now, comprehensive efforts have been made to explore the origin of PSS in the nuclear spectrum since its recognition. In 1977, a great progress was made by Ginocchio who pointed out that PSS is a relativistic symmetry in the Dirac Hamiltonian and is exactly conserved when the scalar and vector potentials satisfying $\Sigma(r) \equiv S(r) + V(r) = 0$ [15]. Meanwhile, he also revealed that the pseudo-orbital angular momentum $\tilde{l}$ is nothing but the orbital angular momentum of the lower component of the Dirac wave function [15], and certain similarities exist in the relativistic single-nucleon wave functions [16]. Later, Meng et al. pointed out a more general condition of $d\Sigma(r)/dr = 0$, which can be approximately satisfied in exotic nuclei with highly diffuse potentials [17,18] and the onset of the pseudospin symmetry to a competition between the pseudo-centrifugal barrier (PCB) and the pseudospin-orbit (PSO) potential. Afterwards, PSS in nuclear spectra have been studied extensively, such as PSS in deformed nuclei [19–26], spin symmetry (SS) in anti-nucleon spectra [27–31], PSS and SS in hypernuclei [32–34], a perturbative interpretation of SS and PSS [31,35–38], and PSS in supersymmetric quantum mechanics [39–42].

Most studies for PSS have been performed in bound states. However, PSS is always broken in bound states according to the conservation condition and there is no bound state in the PSS limit. In contrast, the resonant states provide us with a better platform for studying PSS, which can be obtained both in the PSS limit and in finite-depth potentials. Besides, resonant states play essential roles in exotic nuclei, where valence nucleons can be easily scattered to single-particle resonant states due to pairing correlations, and the couplings between the bound states and the continuum become very important [43–48]. Therefore, the study of

* Corresponding author at: School of Physics, Zhengzhou University, Zhengzhou 450001, China.
  *E-mail address:* ttsunphy@zzu.edu.cn (T.-T. Sun).

PSS in resonant states has attracted increasing attention in recent years and many investigations have been done. PSS and SS in nucleon-nucleus and nucleon-nucleon scattering have been investigated in Refs. [49–52]. In 2004, Zhang et al. confirmed that the lower components of the Dirac wave functions for the resonant PS doublets also have similarity properties [53]. Guo et al. investigated the dependence of pseudospin breaking for the resonant states on the shape of the mean-field potential in a Woods-Saxon form [54–56] as well as on the ratio of neutron and proton numbers [57]. In 2012, a great progress has been achieved by Lu et al. in Ref. [58], where they gave a rigorous justification of PSS in single-particle resonant states and shown that PSS in single-particle resonant states is also exactly conserved under the same conditions as PSS in bound states, i.e., $\Sigma(r) = 0$ or $d\Sigma(r)/dr = 0$ [58,59]. Recently, taking the Green's function method, we studied the conservation and breaking in the resonant states and in the PSS limit, besides the exactly same energy and width between the PS partners, identical density distributions of the lower components have also been identified, which provides a direct evidence that PSS is a relativistic symmetry related with the lower component of Dirac wave functions [60].

Although many investigations have been performed for the PSS in the single-particle bound and resonant states. Still, a uniform description in PSS for bound and resonant states, is absent and highly expected. In this work, following our previous work [60], we will further illustrate the PSS in the single-particle bound and resonant states uniformly in spherical Woods-Saxon potentials by varying potential depth continuously, where PS partners can evolve from resonant states to bound states. The Green's function method [61–67] is employed, which has been confirmed to be one of the most efficient tools for studying the single-particle resonant states due to its advantages such as: the bound and resonant states are treated on the same footing, the energies and widths for all resonances are precisely determined regardless of their widths, the spatial density distributions are properly described, and the great applicability in any potential without any requirement on the potential shapes [68–71].

This paper is organized as follows. The theoretical framework of the Green's function method is briefly presented in Section 2. Section 3 is devoted to the discussion of the numerical results, where the breaking of the PSS in single-particle bound and resonant states are illustrated uniformly by analyzing the splittings in energy, width, and density distributions between the PS partners. Finally, a summary is given in Section 4.

## 2. Theoretical framework

In a relativistic description, nucleons are Dirac spinors moving in a mean-field potential with an attractive scalar potential $S(\mathbf{r})$ and a repulsive vector potential $V(\mathbf{r})$ [72]. The Dirac equation for a nucleon reads

$$[\boldsymbol{\alpha} \cdot \mathbf{p} + V(\mathbf{r}) + \beta(M + S(\mathbf{r}))]\psi_n(\mathbf{r}) = \varepsilon_n \psi_n(\mathbf{r}), \tag{1}$$

where $\boldsymbol{\alpha}$ and $\beta$ are the Dirac matrices and $M$ is the nucleon mass. Based on the Dirac Hamiltonian $\hat{h}(\mathbf{r})$, a relativistic single-particle Green's function $\mathcal{G}(\mathbf{r}, \mathbf{r}'; \varepsilon)$ can be constructed, which obeys

$$[\varepsilon - \hat{h}(\mathbf{r})]\mathcal{G}(\mathbf{r}, \mathbf{r}'; \varepsilon) = \delta(\mathbf{r} - \mathbf{r}'). \tag{2}$$

With a complete set of eigenstates $\psi_n(\mathbf{r})$ and eigenvalues $\varepsilon_n$, the Green's function can be simply represented as

$$\mathcal{G}(\mathbf{r}, \mathbf{r}'; \varepsilon) = \sum_n \frac{\psi_n(\mathbf{r})\psi_n^\dagger(\mathbf{r}')}{\varepsilon - \varepsilon_n}, \tag{3}$$

which is a $2 \times 2$ matrix because of the upper and lower components of the Dirac spinor $\psi_n(\mathbf{r})$. Equation (3) is fully equivalent to Eq. (2).

For a spherical nucleus, the Green's function can be expanded as

$$\mathcal{G}(\mathbf{r}, \mathbf{r}'; \varepsilon) = \sum_{\kappa m} Y^l_{jm}(\theta, \phi) \frac{\mathcal{G}_\kappa(r, r'; \varepsilon)}{rr'} Y^{l*}_{jm}(\theta', \phi'), \tag{4}$$

where $Y^l_{jm}(\theta, \phi)$ is the spin spherical harmonic, $\mathcal{G}_\kappa(r, r'; \varepsilon)$ is the radial Green's function, and the quantum number $\kappa = (-1)^{j+l+1/2}(j + 1/2)$. The Eq. (2) can be reduced as

$$\left[\varepsilon - \begin{pmatrix} \Sigma(r) & -\frac{d}{dr} + \frac{\kappa}{r} \\ \frac{d}{dr} + \frac{\kappa}{r} & \Delta(r) - 2M \end{pmatrix}\right] \mathcal{G}_\kappa(r, r'; \varepsilon) = \delta(r - r')I, \tag{5}$$

where $\Sigma(r) \equiv V(r) + S(r)$, $\Delta(r) = V(r) - S(r)$, and $I$ is a two-dimensional unit matrix. For the construction of the radial Green's function $\mathcal{G}_\kappa(r, r'; \varepsilon)$ and other details for the application of the Green's function method in relativistic mean field, please see Refs. [65,68].

Starting from the first order coupled Eq. (5), we can rewrite two decoupled second order differential ones. Here to study the PSS, we only write down the one for "22" matrix element $\mathcal{G}^{(22)}_\kappa(r, r'; \varepsilon)$ which is related with the lower component of Dirac wave function,

$$\left\{ -\frac{1}{M_-}\frac{d^2}{dr^2} + \frac{1}{M_-^2}\frac{dM_-}{dr}\frac{d}{dr} + M_+ + \frac{1}{M_-}\frac{\tilde{l}(\tilde{l}+1)}{r^2} \right.$$
$$\left. -\frac{1}{M_-^2}\frac{dM_-}{dr}\frac{\kappa}{r} \right\} \mathcal{G}^{(22)}_\kappa(r, r'; \varepsilon) = \delta(r - r'), \tag{6}$$

where $M_+ = \varepsilon + 2M - \Delta(r)$, $M_- = \varepsilon - \Sigma(r)$, and $\tilde{l} = l - \mathrm{sgn}(\kappa)$. From this Schrödinger-like equation, the pseudo centrifugal barrier (PCB) and the pseudo spin-orbit (PSO) potential can be obtained,

$$V_{\mathrm{PCB}}(r) = \frac{1}{M_-}\frac{\tilde{l}(\tilde{l}+1)}{r^2}, \tag{7}$$

$$V_{\mathrm{PSO}}(r) = -\frac{1}{M_-^2}\frac{dM_-}{dr}\frac{\kappa}{r}, \tag{8}$$

the competition of which determines the splitting and conservation of PSS.

Thus, the PSO splitting energy can be obtained as

$$V_{\mathrm{PSO}} = -\int_0^{R_{\mathrm{box}}} dr F^2(r) \frac{1}{M_-^2}\frac{dM_-}{dr}\frac{\kappa}{r}, \tag{9}$$

where $F(r)$ is the lower component of Dirac wave function, and $R_{\mathrm{box}}$ is the space size. In the framework of Green's function method, it is

$$V_{\mathrm{PSO}} = -\int_0^{R_{\mathrm{box}}} dr \left( \frac{1}{2\pi i} \oint_C d\varepsilon \mathcal{G}^{(22)}_\kappa(r, r; \varepsilon) \right) \frac{1}{M_-^2}\frac{dM_-}{dr}\frac{\kappa}{r}. \tag{10}$$

For the mean-field potentials, radial Woods-Saxon like potentials are considered both for $\Sigma(r)$ and $\Delta(r)$,

$$\Sigma(r) = \frac{C}{1 + e^{(r-R)/a}}, \quad \Delta(r) = \frac{D}{1 + e^{(r-R)/a}}, \tag{11}$$

where different depths $C$ of Fermi sea are taken, i.e., from the PSS limit $C = 0$ to finite depths $C = -10, -20, \cdots, -80$ MeV to trace the PS partners continuously. The potential depth for the Dirac sea is fixed as $D = 650$ MeV, the width $R = 7$ fm, and the diffusivity parameter $a = 0.3$ fm following our previous work [60].

## 3. Results and discussion

On the single-particle complex energy plane, as shown in Fig. 1, the bound and resonant states are distributed respectively on the negative real energy axis and in the fourth quadrant. The energy $\varepsilon_n$ is real for bound states while complex for resonant states and in the latter case $\varepsilon_n = E - i\Gamma/2$ with $E$ and $\Gamma$ being the resonant energy and the width respectively. Meanwhile, as shown in Eq. (3), these eigenvalues are also the poles of the Green's function. Thus, in Refs. [69–71] it has been proposed to determine the single-particle energies $\varepsilon_n$ by searching for the poles of the Green's function. In practice, one can do this by calculat-ing the integral function of the Green's function $G_\kappa(\varepsilon)$ for each partial wave $\kappa$ at different energies $\varepsilon$ [71]

$$G_\kappa(\varepsilon) = \int dr \left( |\mathcal{G}_\kappa^{(11)}(r,r;\varepsilon)| + |\mathcal{G}_\kappa^{(22)}(r,r;\varepsilon)| \right), \quad (12)$$

where $|\mathcal{G}_\kappa^{(11)}(r,r;\varepsilon)|$ and $|\mathcal{G}_\kappa^{(22)}(r,r;\varepsilon)|$ are the moduli of the Green's functions respectively for the "11" and "22" matrix elements. This approach has been certified to be highly effective for all resonant states regardless of width [70,71]. To search for the bound and resonant states, Green's functions in a wide energy range are calculated by scanning the single-particle energy $\varepsilon$. For the bound states, the energies $\varepsilon$ are taken along the negative real energy axis. For the resonant states, the energies $\varepsilon$ are complex $\varepsilon = \varepsilon_r + i\varepsilon_i$ which are scanned in the fourth quadrant of the complex energy plane $\varepsilon$, both along the real $\varepsilon_r$ and imaginary energy $\varepsilon_i$ axes.

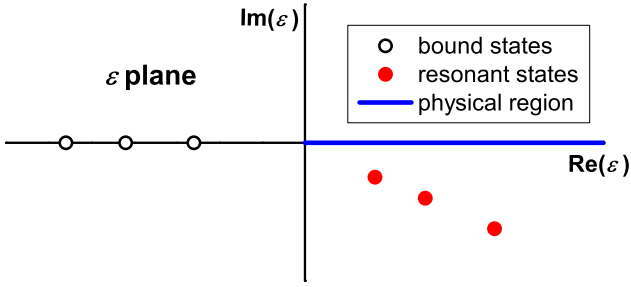

**Fig. 1.** Single-particle complex energy plane with bound states and resonant states.

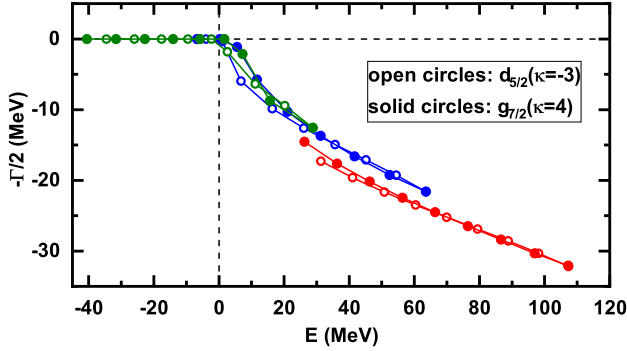

**Fig. 2.** Poles of the Green's function on the complex energy plane in potentials with different depths $C = 0, -10, -20, \cdots, -80$ MeV for the PS partner with the pseudo-orbital angular momentum $\tilde{l} = 3$: $d_{5/2}$ ($\kappa = -3$) and $g_{7/2}$ ($\kappa = 4$). The intruder states are not shown.

In order to study PSS in bound and resonant states uniformly, PS partners are traced continuously from the PSS limit to finite-depth potentials. In Fig. 2, we show the poles of Green's function on the complex energy plane with different potential depths $C = 0, -10, -20, \cdots, -80$ MeV for the PS doublets with pseudo-orbital momentum $\tilde{l} = 3$. Calculations are done with an energy step of 0.1 keV for the integral functions $G_\kappa(\varepsilon)$ in a coordinate space with size $R_{\max} = 20$ fm and a step of $dr = 0.05$ fm. PS partners with pseudospin $\tilde{s} = \pm 1/2$ are denoted as solid and open circles, respectively. Three groups of $\tilde{l} = 3$ PS partners are obtained and the $1\tilde{f}$ ($2d_{5/2}, 1g_{7/2}$), $2\tilde{f}$ ($3d_{5/2}, 2g_{7/2}$), and $3\tilde{f}$ ($4d_{5/2}, 3g_{7/2}$) doublets are denoted in red, blue, and olive, respectively. In the PSS limit, i.e., $C = 0$, the PS partners overlap completely, indicating the exact conservation of PSS. As the potential depth increases, the PS resonance partners move diagonally upward until becoming bound states, distributed along the negative energy real axis. In this process, the breaking of PSS is observed with the energy and width splitting between the PS partners. This provides us an opportunity to give a uniform description of the breaking mechanisms in the PSS for the bound and resonant states.

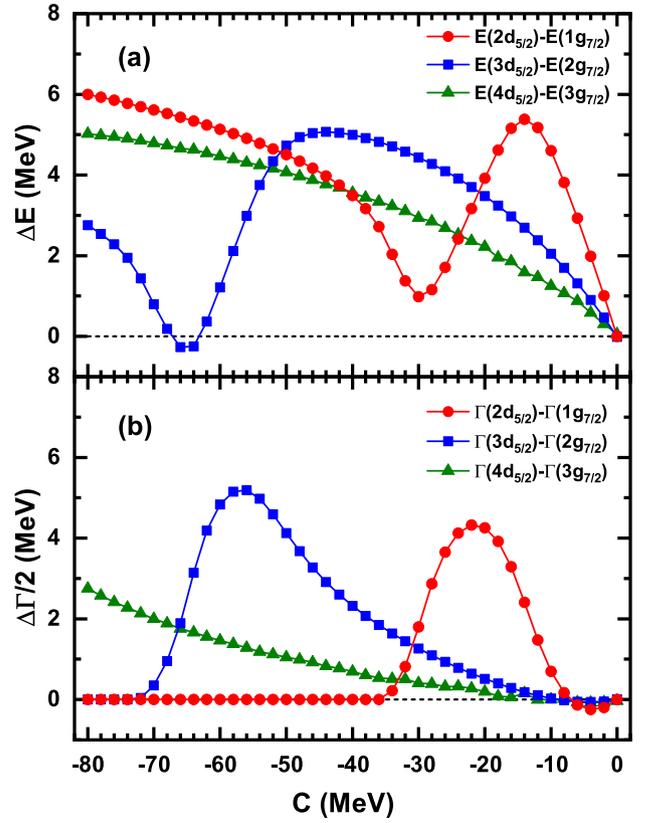

**Fig. 3.** Energy and width splittings between the PS partners with $\tilde{l} = 3$ as a function of the potential depth $C$.

In Fig. 3, the energy and width splittings between $\tilde{l} = 3$ PS partners are plotted as a function of the potential depth $C$. In the following, we will take the PS partner $2\tilde{f}$ ($3d_{5/2}, 2g_{7/2}$) as an example, the evolution of which with the potential depth $C$ covers both the resonant states and the bound states. For the energy splitting shown in Fig. 3(a), starting from the zero potential depth, it first increases and then decreases until encountering the threshold at a critical value of potential depth, $C_0 \approx -65$ MeV, and again the splitting increases. To further illustrate the threshold, we plot in Fig. 4 the relative locations of the PS partners and the corresponding centrifugal barriers $\Sigma(r) + \frac{1}{M_+}\frac{l(l+1)}{r^2}$. In Fig. 4(a), with the increase of the potential depth $C$, the single-particle levels goes down continuously, and gradually become quasi-bound states inside the centrifugal barrier from the high-energy resonante states. Note that the centrifugal barriers vary slightly when the potential depth changes from $-60$ to $-70$ MeV and we only plot that for $C_0 = -65$ MeV. The PS partner $2g_{7/2}$ and $3d_{5/2}$ drop down to lower than the corresponding centrifugal barriers when $C = -60$ and $-70$ MeV, respectively. And the levels of the PS partner obtained with the critical potential depth $C_0$ corresponds to the case that $2g_{7/2}$ has become a quasi-bound state inside the centrifugal barrier while $3d_{5/2}$ is still a high-energy resonant state outside the centrifugal barrier. As shown in Fig. 4(b), the energy difference between the single-particle levels and the barrier of centrifugal potential is plotted, which is positive when levels locate above the barrier while negative when they locate below the barrier. Around the critical potential depth $C_0$, the PS partner $2g_{7/2}$ and $3d_{5/2}$ has the closest value of $|E - E_{\text{barrier}}|$. According to the location of $C_0$, we can roughly divide the morphology of the PS partners into two segments, and when the potential depth $|C| < |C_0|$ the partners are mainly resonance states, and when $|C| > |C_0|$ they display as bound states. This kind of threshold effect is also observed for other PS partners and only the critical values of $C_0$ are different. Besides, for the $2\tilde{f}$ PS doublet, the level inversion occurs around the potential depth $C_0$, where the energy splitting $\Delta E = E(3d_{5/2}) - E(2g_{7/2})$ becomes negative. In many previous studies [33,55,56], the level inversion phenomena have also been found for resonant PS doublets both in spherical and deformed nuclei. But this does not always happen for each PS doublet, such as $1\tilde{f}$ ($2d_{5/2}, 1g_{7/2}$).

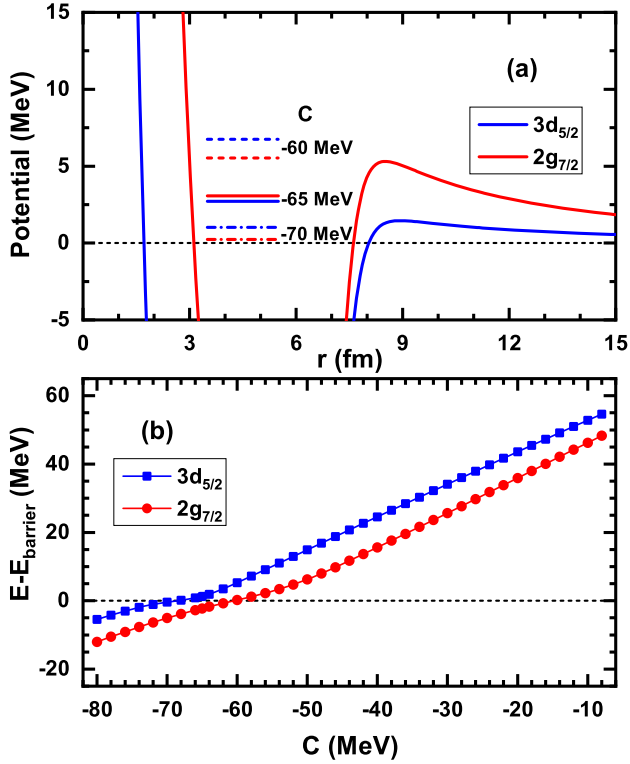

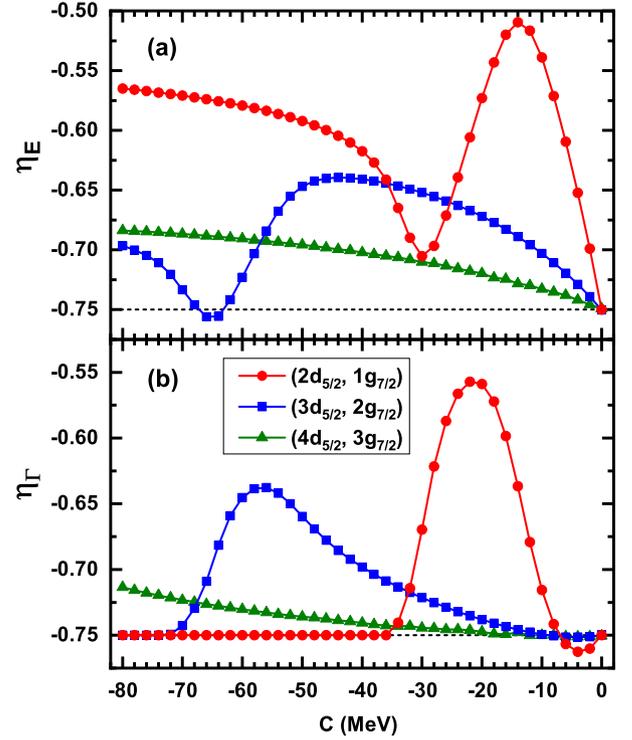

**Fig. 4.** (a) The PS partner $3d_{5/2}$ ($\kappa = -3$) and $2g_{7/2}$ ($\kappa = 4$) and the corresponding mean field potentials $\Sigma(r) + V_{\text{cent}}(r)$ at the critical potential depth $C_0 = -65$ MeV. For comparison, levels obtained with $C = -60, -70$ MeV are also plotted. (b) The energy difference $E - E_{\text{barrier}}$ between the single-particle levels and the barrier of centrifugal potential $V_{\text{cent}}(r)$ for the PS partner $3d_{5/2}$ and $2g_{7/2}$ as a function of potential depths $C$.

**Fig. 5.** The ratio between the PSO potentials for the $\tilde{l} = 3$ PS partner $d_{5/2}$ ($\kappa = -3$) and $g_{7/2}$ ($\kappa = 4$) as a function of the potential depth $C$.

For the width splitting $\Delta\Gamma/2$ plotted in Fig. 3(b), it firstly decreases to a negative value for a shallow potential, e.g. $C \approx -5$ MeV, and then increases rapidly till a maximum around 5 MeV as the potential deepens. Whereafter, the width splitting drops to zero when the PS partner becomes to be bound states. For each PS partner, the width splitting reaches its largest value at some point of potential depth $C$ before the PS partners enter the centrifugal barrier but with the largest slope of the energy splitting $\Delta E$ with $C$. When the potential depth is very shallow, the PS partners especially $4d_{5/2}$ and $3g_{7/2}$ are very high-energy resonant states, in this case, the influences by the centrifugal barrier on the resonant widths are weak. However, when the PS partners approach to the centrifugal barrier especially after they entering the barrier, the resonant widths are reduced greatly and for each PS doublet, the state $d_{5/2}$ owns larger width compared with $g_{7/2}$ due to the lower centrifugal barrier.

To understand the splitting in Fig. 3 and describe the breaking mechanism of PSS in bound and resonant states uniformly, the PSO potential for the PS partners are compared. Due there are some singular points in the term $M_-$, it failed when taking Eq. (9) directly to study the energy and width splittings between the PS partners. Instead, we do this by calculating the ratio $\eta$ of the PSO potentials between the PS partners, and for the ($d_{5/2}, g_{7/2}$) PS partner, we have

$$\eta = \frac{V_{\text{PSO}}(d_{5/2})}{V_{\text{PSO}}(g_{7/2})} = -\frac{3}{4} \frac{M_-^2(g_{7/2})}{M_-^2(d_{5/2})}$$

$$= -\frac{3}{4}\left(\frac{\varepsilon_{g_{7/2}} - \Sigma}{\varepsilon_{d_{5/2}} - \Sigma}\right)^2$$

$$\approx -\frac{3}{4}\left(1 - \frac{\varepsilon_{d_{5/2}} - \varepsilon_{g_{7/2}}}{\varepsilon_{d_{5/2}} - C}\right)^2. \tag{13}$$

In the last step of Eq. (13), an approximation that $\Sigma(r) = C$ is taken considering the very small diffusivity parameter. Note that the value of $\eta$ is mainly decided by the single-particle energy difference $\varepsilon_{d_{5/2}} - \varepsilon_{g_{7/2}}$ and $\eta = -3/4$ correspond to the PSS limit, in which one has $C = 0$ and $\varepsilon_{d_{5/2}} = \varepsilon_{g_{7/2}}$. For the resonant states, we define $\eta_E$ and $\eta_\Gamma$ respectively related with the energy and width splitting as

$$\eta_E = -0.75\left(1 - \frac{\Delta E}{\varepsilon_{d_{5/2}} - C}\right)^2, \tag{14}$$

$$\eta_\Gamma = -0.75\left(1 - \frac{\Delta\Gamma/2}{\varepsilon_{d_{5/2}} - C}\right)^2. \tag{15}$$

In Fig. 5, we plot $\eta_E$ and $\eta_\Gamma$ as functions of potential depth $C$ for the $\tilde{l} = 3$ PS doublets. Almost the same shape for the evolution with the potential depth $C$ is shown as those for the energy and width splittings in Fig. 3, where exactly the same critical potential depths $C_0$ are obtained. All those indicate that the breaking of the PSS both for the resonant and bound states are directly determined by the relative magnitudes of the PSO potentials between the PS partners.

To check the conservation and breaking mechanism of PSS, the similarities of the Dirac wave functions for the PS doublets provide another important way. With the Green's function method, the density distributions in the coordinate space can also be examined by exploring $\rho_\kappa(r,\varepsilon)$ defined at the energy $\varepsilon = E$,

$$\rho_\kappa(r,\varepsilon) \tag{16}$$
$$= -\frac{1}{4\pi r^2}\frac{1}{\pi}\text{Im}\left[\mathcal{G}_\kappa^{(11)}(r,r;E) + \mathcal{G}_\kappa^{(22)}(r,r;E)\right],$$

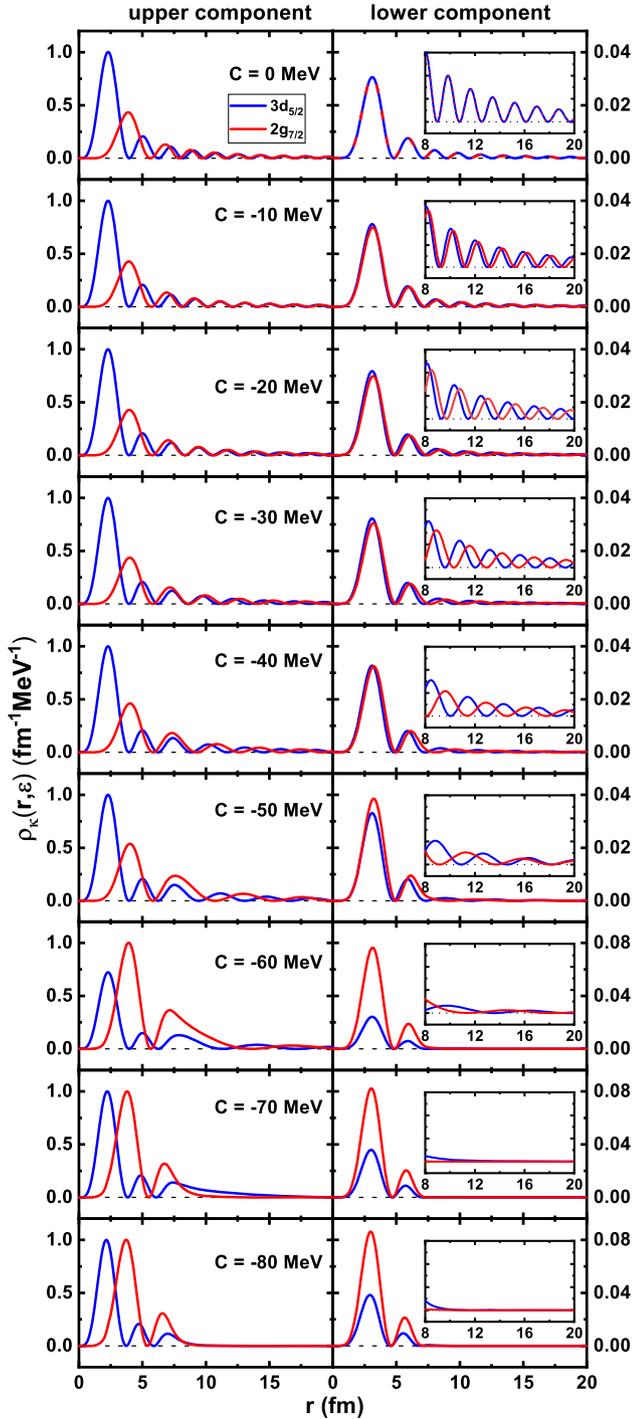

**Fig. 6.** The density distributions $\rho_\kappa(r,\varepsilon)$ for the PS doublets $3d_{5/2}$ and $2g_{7/2}$ for various depths of the potentials, from the PSS limit with $C = 0$ MeV to the cases with $C = -10, -20, \cdots, -80$ MeV. Those plotted in the left and right columns correspond to the upper and lower components of the Dirac wave functions, respectively.

where the terms $\mathcal{G}^{(11)}_\kappa(r,r;E)$ and $\mathcal{G}^{(22)}_\kappa(r,r;E)$ are respectively related to the upper and lower components of the Dirac wave functions. In Fig. 6, taking the PS doublet $(3d_{5/2}, 2g_{7/2})$ as an example, we investigate the evolutions of the density distributions $\rho_\kappa(r,\varepsilon)$ with different potential depths $C$ continuously from the PSS limit with $C = 0$ to finite depth $C = -80$ MeV. The left and right columns are respectively contributed by the upper and lower components of Dirac wave functions. In order to make better comparisons, at each potential depth $C$, we adjust the highest peak of $\rho_\kappa(r,\varepsilon)$ to be 1.0 fm$^{-1}\cdot$MeV$^{-1}$ and ensure that the relative sizes of the different density distributions remain unchanged for the resonant PS partners. In the PSS limit, the density distributions are exactly same for the lower component of PS doublet while they differ a node for the upper components, which is a direct evidence for that PSS is a symmetry of the Dirac Hamiltonian with the pseudo-orbital momentum of the lower component of the Dirac spinor [60]. With the potential depth $C$ deepens to $-10$ MeV, the splitting of PS partner happens, as a result, a slight difference between the phase shifts occurs in the density distributions for the lower components. This kind of phase shift between the PS partners increases as the potential depth $C$ deepens. When the depth of potential further increases and the PS partner $(3d_{5/2}, 2g_{7/2})$ becomes bound states, the oscillation of the density distributions disappear completely and the phase shift disappear. In this case, the difference in the density distributions of the lower component between the PS doublet is mainly manifested in the difference of amplitudes. Note that for the bound PS partners, we adjust the peak of $\rho_\kappa(r,\varepsilon)$ of upper component to be 1.0 fm$^{-1}\cdot$MeV$^{-1}$ both for $3d_{5/2}$ and $2g_{7/2}$ separately and ensure that the relative sizes of the different density distributions remain unchanged.

## 4. Summary

In summary, the PSS in single-particle bound and resonant states are investigated uniformly within a relativistic framework by exploring the poles of the Green's function in spherical Woods-Saxon potentials of different depths. This work is a continuation of our previous work for PSS in resonant states [60], where the conservation and breaking are discussed.

The PS partners are traced continuously as the potential depth increases from zero to finite depths, which evolve from resonant states to bound states and the PSS is broken gradually with energy, width, and density splittings. Specially, the sizes of the energy and width splittings for the resonant and bound states are directly determined by the ratio of the PSO potentials between the PS partners. Obvious threshold effect is observed for the energy splitting at a critical value $C_0$ of potential depth, which corresponds to the case that $2g_{7/2}$ has become a quasi-bound state inside the centrifugal barrier while $3d_{5/2}$ is still a high-energy resonant state outside the centrifugal barrier. The evolution of the density distributions for the PS partners at different potential depths are also examined continuously, the differences of which are manifested in the phase shifts for the resonant states and the amplitudes for bound states.

### Declaration of competing interest

The authors declare that they have no known competing financial interests or personal relationships that could have appeared to influence the work reported in this paper.

### Data availability

Data will be made available on request.

### Acknowledgements

This work was partly supported by the Natural Science Foundation of Henan Province (Grant No. 242300421156), the National Natural Science Foundation of China (No. U2032141 and No. 12375126), the Open Project of Guangxi Key Laboratory of Nuclear Physics and Nuclear Technology (No. NLK2022-02), the Central Government Guidance Funds for Local Scientific and Technological Development, China (No. Guike ZY22096024), and the Fundamental Research Funds for the Central Universities.